\documentclass{ws-procs9x6-cpt22}
\usepackage[caption=false]{subfig}
\usepackage[letterspace=-25]{microtype}
\begin{document}

\newcommand{\refeq}[1]{(\ref{#1})}
\def\etal {{\it et al.}} 


\title{Modular J-PET with Improved o-Ps  
		Detection Efficiency\\ for CPT Tests}

\author{N.\ Chug$^{1,2}$ and A.\ Gajos$^{1,2}$}

\address{$^1$Jagiellonian University,
         Krakow, Poland}
\address{$^2$Center for Theranostics, Jagiellonian University,
Krakow, Poland}
\author{On behalf of the J-PET Collaboration}

\begin{abstract}
  J-PET is a photon detector built of plastic scintillators, which already has been commissioned for CPT studies in the decays of positronium. In the first experiment, J-PET has achieved a sensitivity to CPT violation at a level of 10$^{-4}$, and now it aims to reach a level of 10$^{-5}$. This will be done by enhancing the three-photon registration efficiency for ortho-positronium decays using a new layer of densely packed plastic scintillators termed Modular J-PET. We present the simulation studies performed for different experimental detection setups to be used for the next CPT test with the Modular J-PET detector.

\end{abstract}

\bodymatter

\section{Introduction}
Discrete-symmetry tests in positronium decays can be done by studying certain non-vanishing angular-correlation operators  odd under particular symmetries.\cite{sozzi_iop} For this test we are interested in the CPT-violation-sensitive operator given by $\vec{S}\cdot (\vec{k_1}\times\vec{k_2})$, which is the angle between the spin and the orientation of the annihilation plane of the ortho-positronium (o-Ps) atom.\cite{arbic} J-PET, conceived as a tomography device, allows for exclusive registration of a broad range of kinematical configurations of three-photon annihilations with large geometrical acceptance and high angular resolution.\cite{operator}

\section{Towards improving the sensitivity of CPT test}
A measurement of the CPT-odd operator $\vec{S}\cdot(\vec{k_1}\times\vec{k_2})$ was done with J-PET consisting of 192 plastic scintillator strips arranged concentrically in three layers with PMT readouts on both ends of each strip.\cite{detector}
Data is collected in a trigger-less mode at four different thresholds applied to photomultiplier signals.\cite{ftab} The first CPT test with J-PET has reached sensitivities better than the best known previous experimental result by a factor of three.\cite{gammasphere,cpt_alek}
Now, the J-PET detector is being upgraded with an additional layer consisting of 24 modules of densely packed plastic scintillators with 13 scintillators in each module and silicon photomultiplier readouts. It will result in improving the time resolution and enhancing the angular acceptance of the detector. 
Another possible improvement is to increase the formation probability of o-Ps atoms that is achieved by replacing the cylindrical annihilation chamber (used in the previous experiment) with a spherical vacuum chamber.\cite{cpt_alek, alek}

\section{Future CPT test with Modular J-PET}
\begin{figure}[h!]
 \centering
    \subfloat[]{\includegraphics[height=2.2cm]{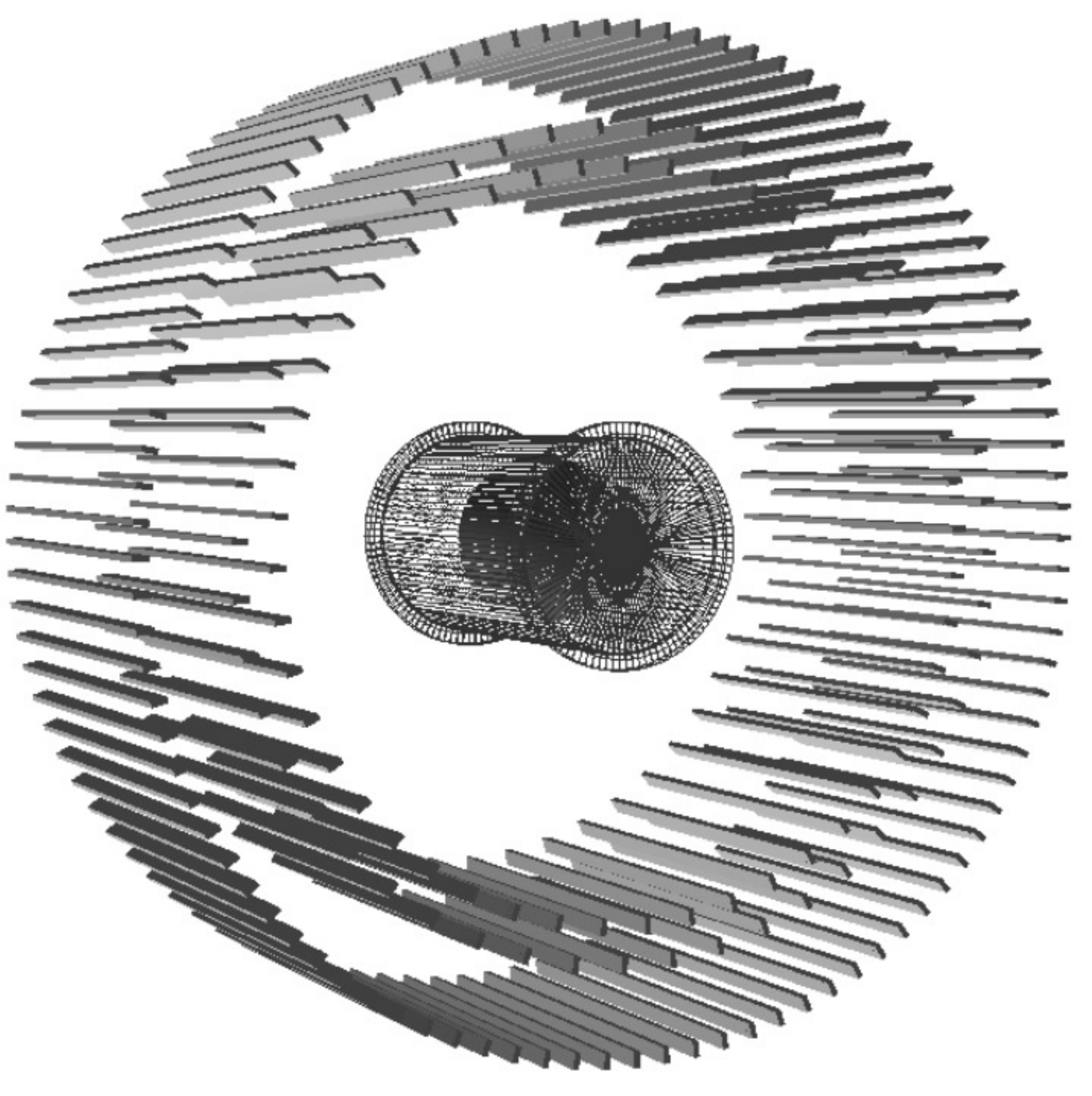}}\hspace{1em}
    \subfloat[]{\includegraphics[height=2.2cm]{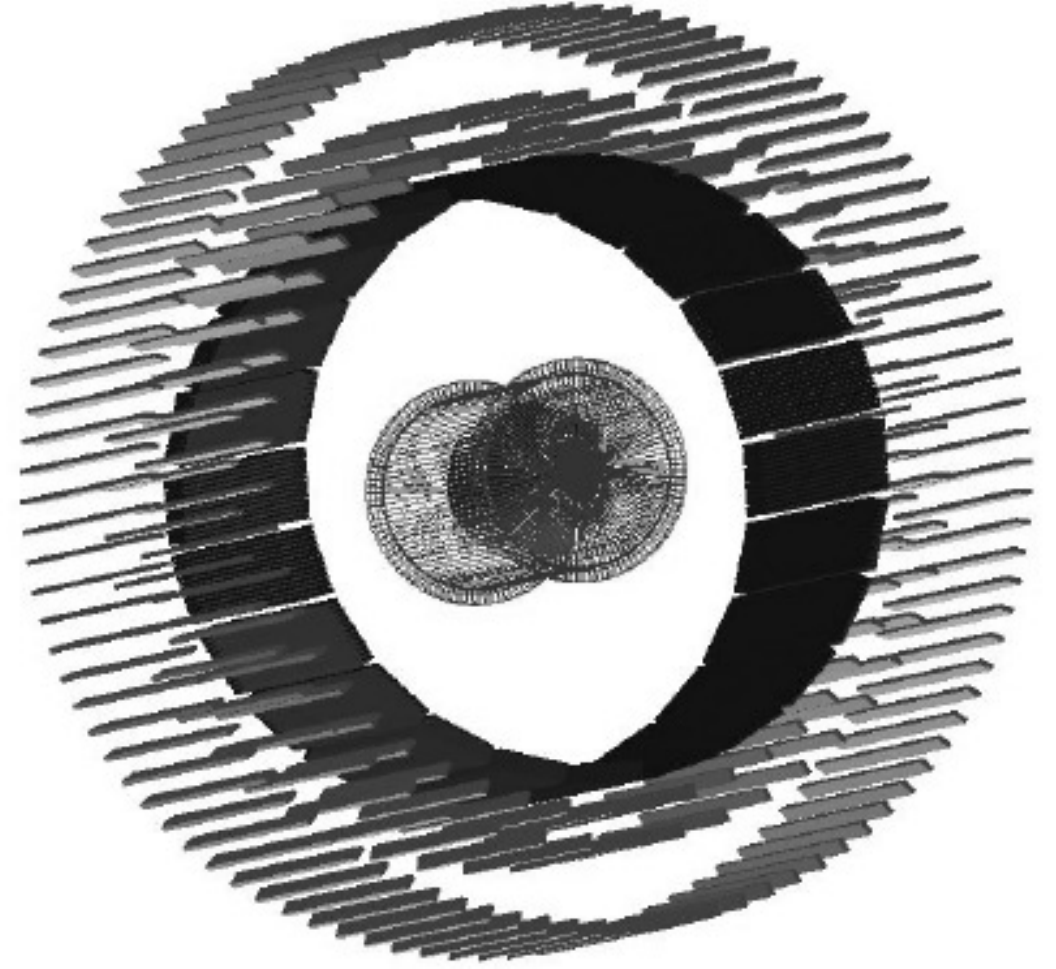}}\hspace{1em}
    \subfloat[]{\includegraphics[height=2.2cm]{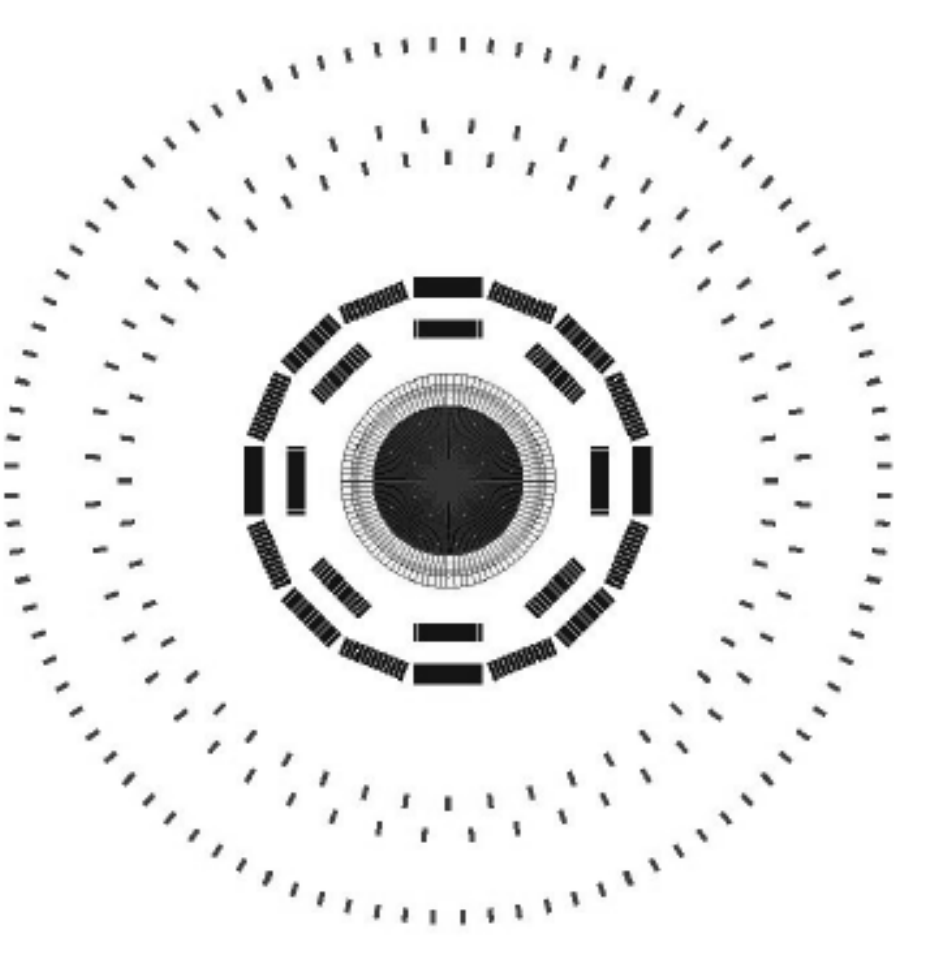}}\hspace{1em}
    \subfloat[]{\includegraphics[height=2.2cm]{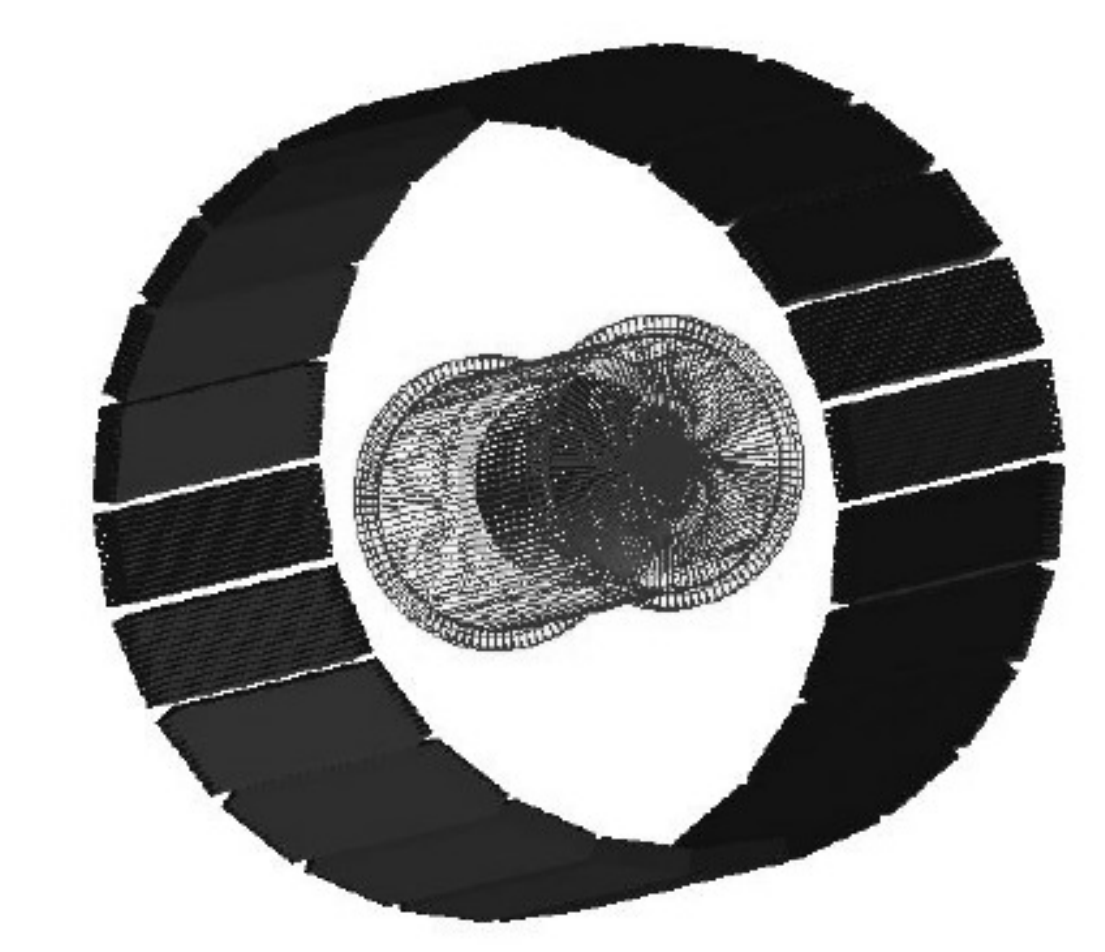}}
    \caption{Cross section of the Modular J-PET detector with spherical annihilation chamber at the center for CPT-symmetry tests with three different  configurations consisting of (a) 3-layer J-PET (already in use), (b) 3-layer J-PET combined with 24 Modular J-PET, (c) 5-layer setup with 16 and 8 modules of Modular J-PET along with 3-layer J-PET,  and (d) 24-Modular digital J-PET used as a stand-alone detector.}  
    \label{fig:configurations}
\end{figure}

Modular J-PET is a compact detector with 24 modules of plastic scintillators, where the modules can be arranged in a single layer or be used as a multi-layer system. We have performed  MC simulations of different geometrical configurations of Modular J-PET and the spherical annihilation chamber. Studies are done to choose the best geometrical configuration to evaluate the measurement time and conditions needed to reach the sensitivity of 10$^{-5}$ to CPT violation either with the originally-proposed combined setup or with a single-layer or multi-layer digital J-PET setup as shown in Fig.~\ref{fig:configurations}. The relative gain in efficiency of the registration of three-photon annihilations of o-PS in different configurations with Modular J-PET with respect to the 3-layer J-PET (Fig.~\ref{fig:configurations}(a)) used in the previous experiments, as obtained with MC simulations, is given in Table~\ref{table:1}.

\begin{table}[h!]
\tbl{Relative efficiency for registration of o-Ps${\rightarrow}3{\gamma}$ events; fraction of secondary Compton-scattered events recorded in different detector geometries (MC simulation).}
{\begin{tabular}{@{}cccc@{}}\toprule
Detector geometry & Relative Efficiency  & Fraction    \\
 & (w.r.t.\ 3 layer J-PET) & (secondary scatterings) \\\colrule
 (a) 3-layer J-PET   & 1 & 10\%  \\[0.01ex]
 (b) 3-layer J-PET+ 24 Modules   & 28 & 37\% \\[0.01ex]
 (c) 3-layer J-PET+ 16+8 Modules   &  112 & 28\% \\[0.01ex]
 (d) 24 Modular J-PET & 17 & 7\% \\\botrule
\end{tabular}}
\label{table:1}
\end{table}

\section{Conclusions and perspectives}
J-PET has already started test measurements with the spherical annihilation chamber and the 3-layer detector setup.\cite{panic} From our MC simulations, we conclude to use the stand-alone digital J-PET detector with spherical annihilation chamber \textit{(d)} for the next CPT test. Although its registration efficiency is less than the multi-layer setups \textit{(b)} and \textit{(c)}, it allows easier data acquisition, detector systematics, and dealing with secondary Compton scatterings compared to the combined setup. The high fraction of secondary Compton-scattered events in the multi-layer detector setup as given in Table~\ref{table:1} would result in an increase in background contributions. With an efficiency as shown in Table~\ref{table:1}, J-PET would be able to search for the CPT violation at the precision level of 10$^{-5}$, and an efficiency gain of a factor by 17 is sufficient to reach the required precision in four months of data taking.

\section*{Acknowledgments}
This work was supported by the Foundation for Polish Science through TEAM/2017-4/39, NCN of Poland through 2019/35/B/ST2/03562, Jagiellonian University MNS grant no.\ 2021-N17/MNW/000013, and SciMat and qLife Priority Research Areas budget under the program Excellence Initiative - Research University at the Jagiellonian University.

\end{document}